\documentclass{article}
\usepackage{spconf,amsmath,graphicx,hyperref}
\usepackage{mathrsfs}
\usepackage{amssymb}
\usepackage{enumitem}
\usepackage{amsmath}
\usepackage{algorithm}
\usepackage{algpseudocode}
\usepackage{graphicx}   

\title{TCDA: Robust 2D-DOA Estimation for Defective L-Shaped Arrays}
\name{Wenlong Wang$^{1 \dagger}$, Tianyang Zhang$^{2 \dagger}$, Tailun Dong$^3$ and Lei Zhang$^{1 \star}$
      \thanks{$^{\dagger}$These authors contributed equally to this work.}
      \thanks{$^{\star}$Corresponding author: Lei Zhang (email: zhang.lei@tsinghua.edu.cn).}
      \thanks{This work was supported by the National High Technology Research and Development Program of China under Grant 2021YFB2900403.}
}
\address{\textsuperscript{1} School of Integrated Circuits,
Tsinghua University,
China\\
\textsuperscript{2} School of Information and Communication Engineering,\\
University of Electronic Science and Technology of China,
China\\
\textsuperscript{3} School of Communications and Information Engineering \& School of Artificial Intelligence,\\
Xi’an University of Posts and Telecommunications,
China
}
\begin{document}
%
\maketitle
\begin{abstract}
While tensor-based methods excel at Direction-of-Arrival (DOA) estimation, their performance degrades severely with faulty or sparse arrays that violate the required manifold structure. To address this challenge, we propose Tensor Completion for Defective Arrays (TCDA) , a robust algorithm that reformulates the physical imperfection problem as a data recovery task within a virtual tensor space. We present a detailed derivation for constructing an incomplete third-order Parallel Factor Analysis (PARAFAC) tensor from the faulty array signals via subarray partitioning, cross-correlation, and dimensional reshaping. Leveraging the tensor's inherent low-rank structure, an Alternating Least Squares (ALS)-based algorithm directly recovers the factor matrices embedding the DOA parameters from the incomplete observations. This approach provides a software-defined 'self-healing' capability, demonstrating exceptional robustness against random element failures without requiring additional processing steps for DOA estimation.
\end{abstract}
\section{Introduction}

DOA estimation is a fundamental task in radar, sonar, and wireless communications~\cite{Li2007, Haimovich2008}. L-shaped arrays are common for 2D DOA estimation~\cite{Hua1991}, where tensor-based methods like PARAFAC offer automatic parameter pairing and increased degrees of freedom (DOF)~\cite{Harshman1970, Pal2010}. However, these methods rely on a precise array manifold and suffer catastrophic performance degradation from sensor failures that corrupt the array's Vandermonde structure. Conventional solutions like hardware maintenance are costly, while calibration algorithms are often complex and suboptimal.

This paper introduces TCDA, a novel software framework that treats a physically faulty array as an incomplete sampling of an ideal data structure~\cite{Kolda2009}. Unlike Compressed Sensing (CS) strategies that flatten array data into vectors, potentially severing inherent multidimensional dependencies~\cite{Donoho2006}, TCDA operates directly on the high-order tensor structure. This preservation of spatiotemporal correlations is conceptually advantageous for handling array defects, although a detailed comparison with vector-based sparse recovery is beyond the scope of this initial work. While related to virtual array concepts~\cite{Hoctor1990, Dong2017}, our key distinction is a process starting from raw signals. We construct an incomplete third-order tensor via subarray partitioning, cross-correlation, and reshaping. The tensor's ideal low-rank PARAFAC structure ensures identifiability~\cite{Jiang2001}, reformulating the DOA problem as a weighted PARAFAC decomposition with missing data~\cite{Tomasi2006}. An ALS-based algorithm then "interpolates" the missing data to directly recover the factor matrices containing the paired 2D DOA information~\cite{Wang2025}.

This data "healing" process directly yields the factor matrices for DOA estimation, enhancing efficiency and accuracy~\cite{Cichocki2015}. Our contributions are: a) we model the incomplete target tensor from faulty array signals; b) we formalize DOA estimation for faulty arrays as a weighted PARAFAC task; and c) we derive the ALS-based TCDA algorithm, demonstrating its robust, high-accuracy performance on arrays with substantial faults~\cite{Nion2009}.

\section{Signal Model and Problem Formulation}
\label{sec:format}
\subsection{Ideal Array Signal and Data Tensors}
Consider an L-shaped array consisting of two orthogonal M-element Uniform Linear Arrays (ULAs) on the x- and z-axes. $K$ uncorrelated far-field narrowband signals impinge from directions $\{(\phi_k, \theta_k)\}$. We partition the x-ULA into $N_x$ overlapping subarrays of size $L_{x1}$ and the z-ULA into $N_z$ subarrays of size $L_{z1}$.

For the x-ULA, the received signal can be structured into a third-order tensor $\mathcal{X} \in \mathbb{C}^{L_{x1} \times N_x \times N}$ ($N$ is the number of snapshots), whose noiseless part can be expressed as:
\begin{equation}
\mathcal{X}_{ideal} = \sum_{k=1}^K \mathbf{a}_{xk}^{(1)} \circ \mathbf{b}_{xk} \circ \mathbf{s}_k, 
\end{equation}
where $\circ$ denotes the outer product and $\mathbf{s}_k \in \mathbb{C}^{N \times 1}$ is the waveform vector of the $k$-th signal. The factor vectors are constructed based on the term $\Theta_k = e^{j \pi \cos \theta_k}$, where $\mathbf{a}_{xk}^{(1)} = [\Theta_k, \Theta_k^2, \ldots, \Theta_k^{L_{x1}}]^T \in \mathbb{C}^{L_{x1} \times 1}$ is the steering vector of the first subarray, and $\mathbf{b}_{xk} = [1, \Theta_k, \ldots, \Theta_k^{N_x-1}]^T \in \mathbb{C}^{N_x \times 1}$ contains the phase relationship between subarrays.

Similarly, for the z-ULA, we have the data tensor $\mathcal{Z} \in \mathbb{C}^{L_{z1} \times N_z \times N}$:
\begin{equation}
\mathcal{Z}_{ideal} = \sum_{k=1}^K \mathbf{a}_{zk}^{(1)} \circ \mathbf{b}_{zk} \circ \mathbf{s}_k, 
\end{equation}
where $\mathbf{a}_{zk}^{(1)} = [1, \Phi_k, \dots, \Phi_k^{L_{z1}-1}]^T$, $\mathbf{b}_{zk} = [1, \Phi_k, \dots,$ $\Phi_k^{N_z-1}]^T$, and $\Phi_k = e^{j \pi \cos \phi_k}$.

\subsection{Virtual Array Construction and Target Tensor}\label{sec:2.2}
We compute the cross-correlation between $\mathcal{X}$ and $\mathcal{Z}$ along the snapshot dimension, yielding a fourth-order tensor $\mathcal{R}_{xz} \in \mathbb{C}^{L_{x1} \times N_x \times L_{z1} \times N_z}$:
\begin{equation}
\begin{aligned}
\mathcal{R}_{xz} &= \mathbb{E}\{\mathcal{X}(:,:,n) \circ \mathcal{Z}(:,:,n)^*\} \\
&= \sum_{k=1}^K \sigma_k^2 \mathbf{a}_{xk}^{(1)} \circ \mathbf{b}_{xk} \circ \mathbf{a}_{zk}^{(1)*} \circ \mathbf{b}_{zk}^*.
\end{aligned}
\end{equation}

Leveraging the conjugate symmetry property of the ULA manifold, we obtain $\bar{\mathcal{R}}_{xz}^*$ by reversing the dimensions and conjugating $\mathcal{R}_{xz}$. We then concatenate these two tensors along a new fifth dimension to form an augmented five-order tensor $\mathcal{R} \in \mathbb{C}^{L_{x1} \times N_x \times L_{z1} \times N_z \times 2}$:
\begin{equation}
\mathcal{R} = \sum_{k=1}^K \sigma_k^2 \mathbf{a}_{xk}^{(1)} \circ \mathbf{b}_{xk} \circ \mathbf{a}_{zk}^{(1)*} \circ \mathbf{b}_{zk}^* \circ \mathbf{u}_k, 
\end{equation}
where $\mathbf{u}_k = [1, \Theta_k^{-M-1} \Phi_k^{M-1}]^T$.

As shown in~\cite{Rao2018}, to isolate the angular parameters for estimation, the tensor $\mathcal{R}$ can be effectively rearranged into another five-order tensor $\mathcal{Q}$:

\begin{equation}
\mathcal{Q} = \sum_{k=1}^K \mathbf{a}_{xk}^{(0)} \circ \mathbf{c}_k \circ \mathbf{a}_{zk}^{(0)*} \circ \mathbf{e}_k \circ \mathbf{p}_k',
\end{equation}
where $\mathbf{p}_k' = \mathbf{p}_k \sigma_k^2$ is the equivalent signal, and the newly introduced factor vectors are defined as $\mathbf{c}_k = [1, \Theta_k]^T$, $\mathbf{e}_k = [1, \Phi_k^{-1}]^T$, $\mathbf{a}_{xk}^{(0)} = [\Theta_k, \ldots, \Theta_k^{L_{x1}-1}]^T$, and $\mathbf{a}_{zk}^{(0)*} = [1, \Phi_k^{-1}, \ldots, \Phi_k^{-L_{z1}+2}]^T$.
The equivalent signal vector $\mathbf{p}_k$ is formed by the Kronecker product of the factor vectors from (4) corresponding to the dimensions that will be merged:
\begin{equation}
\mathbf{p}_k = \mathbf{u}_k \otimes \mathbf{b}_{zk}^* \otimes \mathbf{b}_{xk}.
\end{equation}
       
Finally, we employ the generalized tensorization operation to merge the dimensions of the five-order tensor $\mathcal{Q}$, forming the final third-order target tensor $\mathcal{T}_{ideal}$:
\begin{equation}
\label{equ:7}
\mathcal{T}_{ideal} \triangleq \mathcal{Q}_{\{1,3\}\{2,4\}\{5\}} = \sum_{k=1}^K \mathbf{g}_k \circ \mathbf{h}_k \circ \mathbf{p}_k'. 
\end{equation}

This operation reshapes the tensor $\mathcal{Q}$ into a third-order tensor, $\mathcal{T}_{ideal} \in \mathbb{C}^{L_{x0}L_{z0} \times 4 \times 2N_xN_z}$, by merging its dimensions $\{1,3\}$, $\{2,4\}$, and $\{5\}$. The resulting tensor has a perfect rank-K PARAFAC structure, with factor vectors defined as the virtual array response $\mathbf{g}_k = \mathbf{a}_{zk}^{(0)*} \otimes \mathbf{a}_{xk}^{(0)}$, the angle information vector $\mathbf{h}_k = \mathbf{e}_k \otimes \mathbf{c}_k = [1, \Theta_k, \Phi_k^{-1}, \Theta_k\Phi_k^{-1}]^T$, and the equivalent signal $\mathbf{p}_k'$. The structure of the vector $\mathbf{h}_k$ is critical for the subsequent DOA estimation.


\subsection{Defective Array Model and Final Problem Formalization}
We now introduce array defects, which are modeled using binary mask vectors $\mathbf{m}_x, \mathbf{m}_z \in \{0, 1\}^M$, where '1' denotes a functional element and '0' a failed one. The actually observed signals at the array aperture are:
\begin{equation}
\mathbf{x}_{obs}(t) = \text{diag}(\mathbf{m}_x) \cdot \mathbf{x}_{ideal}(t) + \mathbf{w}_x(t),
\end{equation}
\begin{equation}
\mathbf{z}_{obs}(t) = \text{diag}(\mathbf{m}_z) \cdot \mathbf{z}_{ideal}(t) + \mathbf{w}_z(t),
\end{equation}
where $\mathbf{w}_x(t)$ and $\mathbf{w}_z(t)$ denote the additive noise vectors. Crucially, a physical failure at the $p$-th sensor simultaneously corrupts all subarray entries satisfying the index relationship $l+n-1=p$. This mechanism induces a structured missing pattern characterized by the loss of entire fibers, which inherently challenges the incoherence conditions typically required for robust recovery.

To determine the reliability of data without prior knowledge of $\mathbf{m}_x$ and $\mathbf{m}_z$, we employ a blind detection mechanism. Let $\mathcal{E}$ be the amplitude tensor with entries $[\mathcal{E}]_{ijk} = |[\mathcal{T}_{obs}]_{ijk}|$. The binary mask tensor $\mathcal{W_T}$ is determined via an adaptive thresholding scheme:
\begin{equation}
[\mathcal{W}_{\mathcal{T}}]_{ijk} = \mathbb{I}\left( [\mathcal{E}]_{ijk} > \eta_{\text{th}} \right),
\end{equation}
where $\mathbb{I}(\cdot)$ is the indicator function which equals 1 if the condition is true and 0 otherwise, and $\eta_{\text{th}}$ is the detection threshold. This threshold is defined based on robust energy statistics as:
\begin{equation}
\label{eq:threshold_def}
\eta_{\text{th}} = \frac{1}{\gamma} \cdot \text{median}\left( \text{vec}(\mathcal{E}) \right) + \epsilon,
\end{equation}
where $\text{vec}(\mathcal{E})$ denotes the vectorized form of the amplitude tensor, $\gamma$ is a sensitivity scaling factor, and $\epsilon$ is a small regularization constant ensuring numerical stability.

This process yields the final mask $\mathcal{W_T} \in \{0, 1\}^{L_{x0}L_{z0} \times 4 \times 2N_xN_z}$, where $w_{ijk}=1$ indicates reliable data. Consequently, our observation model is accurately represented as:
\begin{equation}
\mathcal{T}_{obs} = \mathcal{W_T} \times \mathcal{T}_{ideal} + \mathcal{N_T},
\end{equation}
where $\mathcal{T}_{obs}$ is the incomplete tensor computed from the defective array data, $\times$ denotes the Hadamard product, and $\mathcal{N_T}$ is the noise tensor.

The core problem of this paper is formalized as: given the incomplete, noisy third-order tensor $\mathcal{T}_{obs}$ and the estimated mask $\mathcal{W_T}$, directly recover the factor matrices $\mathbf{G}, \mathbf{H}, \mathbf{P}$ of the ideal tensor $\mathcal{T}_{ideal}$.

\section{TCDA FRAMEWORK: WEIGHTED PARAFAC DECOMPOSITION}
We leverage the low-rank property of $\mathcal{T}_{ideal}$ to complete the missing data and estimate the factor matrices by solving a weighted PARAFAC decomposition problem.

\subsection{Mathematical Formulation}
Our goal is to find a rank-$R$ (where $R$ is the estimate of the number of sources, $K$) third-order tensor $\hat{\mathcal{T}} = [\![\mathbf{G}, \mathbf{H}, \mathbf{P}]\!]$ that minimizes the weighted error with respect to the observed data $\mathcal{T}_{obs}$ at the known entries. The optimization problem is formulated as:
\begin{equation}
\min_{\mathbf{G},\mathbf{H},\mathbf{P}} \| \mathcal{W_T} \times ([\![\mathbf{G}, \mathbf{H}, \mathbf{P}]\!] - \mathcal{T}_{obs}) \|_F^2,
\end{equation}
This is a standard weighted PARAFAC decomposition problem for data with missing entries, which can be solved efficiently using ALS.

\subsection{ Algorithm Derivation via ALS}
We alternately optimize one factor matrix while fixing the other two. Updating the factor matrix $\mathbf{G}$ while keeping $\mathbf{H}$ and $\mathbf{P}$ fixed, the mode-1 unfolding (matricization) of $\hat{\mathcal{T}}$ is given by:
\begin{equation}
\hat{\mathbf{T}}_{(1)}=\mathbf{G}\,(\mathbf{P}\odot\mathbf{H})^{\top},
\end{equation}
where $\odot$ denotes the Khatri--Rao product. The cost function then becomes:
\begin{equation}
\min_{\mathbf{G}} \| \mathbf{W}_{\mathcal{T},(1)} \odot (\mathbf{T}_{obs,(1)} - \mathbf{G} (\mathbf{P} \odot \mathbf{H})^T) \|_F^2,
\end{equation}
This problem can be solved independently for each row $\mathbf{g}_i^T$ of $\mathbf{G}$. For the $i$-th row:
\begin{equation}
\min_{\mathbf{g}_i^T} \sum_{j,k \text{ s.t. } w_{ijk}=1} |t_{obs,ijk} - \mathbf{g}_i^T (\mathbf{p}_j \odot \mathbf{h}_k)|^2,
\end{equation}
where $\mathbf{p}_j^T, \mathbf{h}_k^T$ are row vectors of $\mathbf{P}$ and $\mathbf{H}$, respectively. This is a standard weighted least-squares problem with solution given by the normal equations:
\begin{equation}
\begin{aligned}
&\mathbf{g}_i^T \left( \sum_{(j,k) \in \Phi_i} (\mathbf{p}_j \odot \mathbf{h}_k) (\mathbf{p}_j \odot \mathbf{h}_k)^H \right) \\
&= \sum_{(j,k) \in \Phi_i} t_{obs,ijk} (\mathbf{p}_j \odot \mathbf{h}_k)^H,  
\end{aligned}
\end{equation}
where $\Phi_i$ is the set of index pairs $(j,k)$ satisfying $w_{ijk}=1$. Letting $\mathbf{Z} = \mathbf{P} \odot \mathbf{H}$, the above can be written more compactly:

\begin{equation}
\mathbf{g}_i^T \leftarrow \left( \mathbf{T}_{obs,i,:} \mathbf{W}_{\mathcal{T},i} \mathbf{Z}^* \right) \left( \mathbf{Z}^T \mathbf{W}_{\mathcal{T},i} \mathbf{Z}^* \right)^{-1},
\end{equation}
where $\mathbf{W}_{\mathcal{T},i}$ is a diagonal matrix with diagonal elements from the $i$-th slice of the mask.

The updates for $\mathbf{H}$ and $\mathbf{P}$ follow symmetrically. For example, when updating $\mathbf{H}$, we fix $\mathbf{G}$ and $\mathbf{P}$ and solve the weighted least-squares problem with respect to $\mathbf{H}$.

\subsection{DOA Estimation from Tensor Completion Results}

\begin{algorithm}
\caption{TCDA Based on ALS}
\label{alg:TISAR}
\begin{algorithmic}[1]
\State \textbf{Input:} Incomplete third-order tensor $\mathcal{T}_{\text{obs}}$, mask tensor $\mathcal{W_T}$, desired rank $R$.
\State \textbf{Initialize:} Randomly initialize factor matrices $\mathbf{G}^{(0)}, \mathbf{H}^{(0)}, \mathbf{P}^{(0)}$.
\Repeat
    \State \textbf{Update G:} Fix $\mathbf{H}^{(t)}, \mathbf{P}^{(t)}$, solve for each row update according to (6) to obtain $\mathbf{G}^{(t+1)}$.
    \State \textbf{Update H:} Fix $\mathbf{G}^{(t+1)}, \mathbf{P}^{(t)}$, symmetrically solve to update $\mathbf{H}^{(t+1)}$.
    \State \textbf{Update P:} Fix $\mathbf{G}^{(t+1)}, \mathbf{H}^{(t+1)}$, symmetrically solve to update $\mathbf{P}^{(t+1)}$.
\Until{The fitted DOA error is less than the threshold of $1^\circ$.}
\State \textbf{Output:} Estimated factor matrices $\hat{\mathbf{G}}, \hat{\mathbf{H}}, \hat{\mathbf{P}}$.
\end{algorithmic}
\end{algorithm}

The direct output of Algorithm \ref{alg:TISAR} is the estimated factor matrices $\{\hat{\mathbf{G}}, \hat{\mathbf{H}}, \hat{\mathbf{P}}\}$. No additional PARAFAC decomposition is required. The DOA information is directly contained in factor matrix $\hat{\mathbf{H}} \in \mathbb{C}^{4 \times R}$.

We can directly extract angular information from the elements of the $k$-th column $\hat{\mathbf{h}}_k$ of $\hat{\mathbf{H}}$. The azimuth-related term $\Theta_k = e^{j\pi\cos\theta_k}$ can be estimated as $\hat{\Theta}_k = \hat{h}_{k,2} / \hat{h}_{k,1}$, while the elevation-related term $\Phi_k = e^{j\pi\cos\phi_k}$ is obtained through $\hat{\Phi}_k = (\hat{h}_{k,3} / \hat{h}_{k,1})^{-1}$. Subsequently, the actual DOA angles are recovered from these phase terms through:
\begin{equation}
\hat{\theta}_k = \arccos\left(\frac{\text{angle}(\hat{\Theta}_k)}{\pi}\right), \hat{\phi}_k = \arccos\left(\frac{\text{angle}(\hat{\Phi}_k)}{\pi}\right)
\end{equation}

This extraction process is performed for each column of $\hat{\mathbf{H}}$ to obtain all automatically paired 2D DOA estimates $\{(\hat{\phi}_k, \hat{\theta}_k)\}_{k=1}^R$.







\section{SIMULATION RESULTS}

\begin{figure}[t!]
  \centering
  \begin{minipage}[b]{0.48\linewidth}
    \centering
    \includegraphics[width=\linewidth]{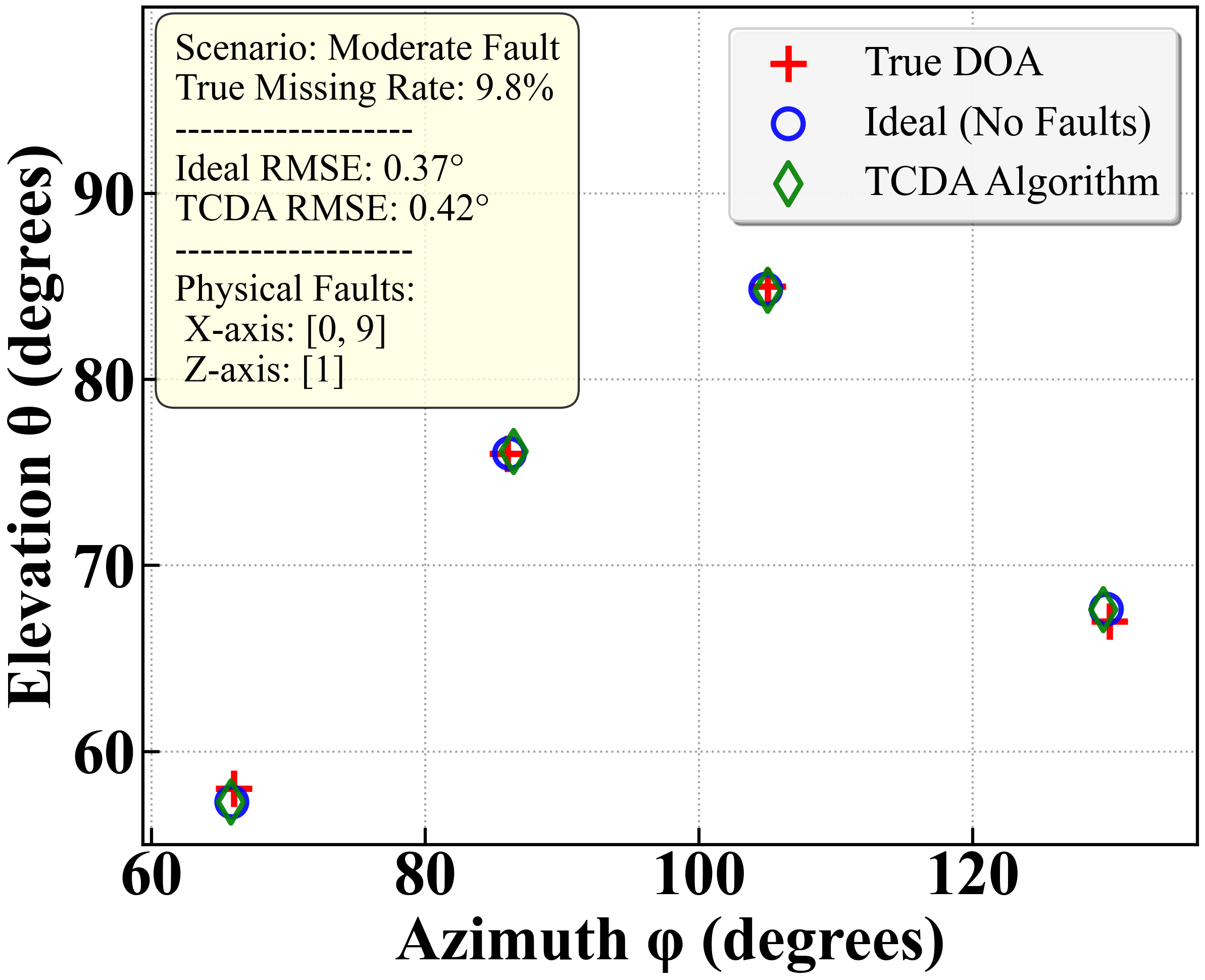}
    \centerline{(a) Moderate}\medskip
  \end{minipage}
  \hfill
  \begin{minipage}[b]{0.48\linewidth}
    \centering
    \includegraphics[width=\linewidth]{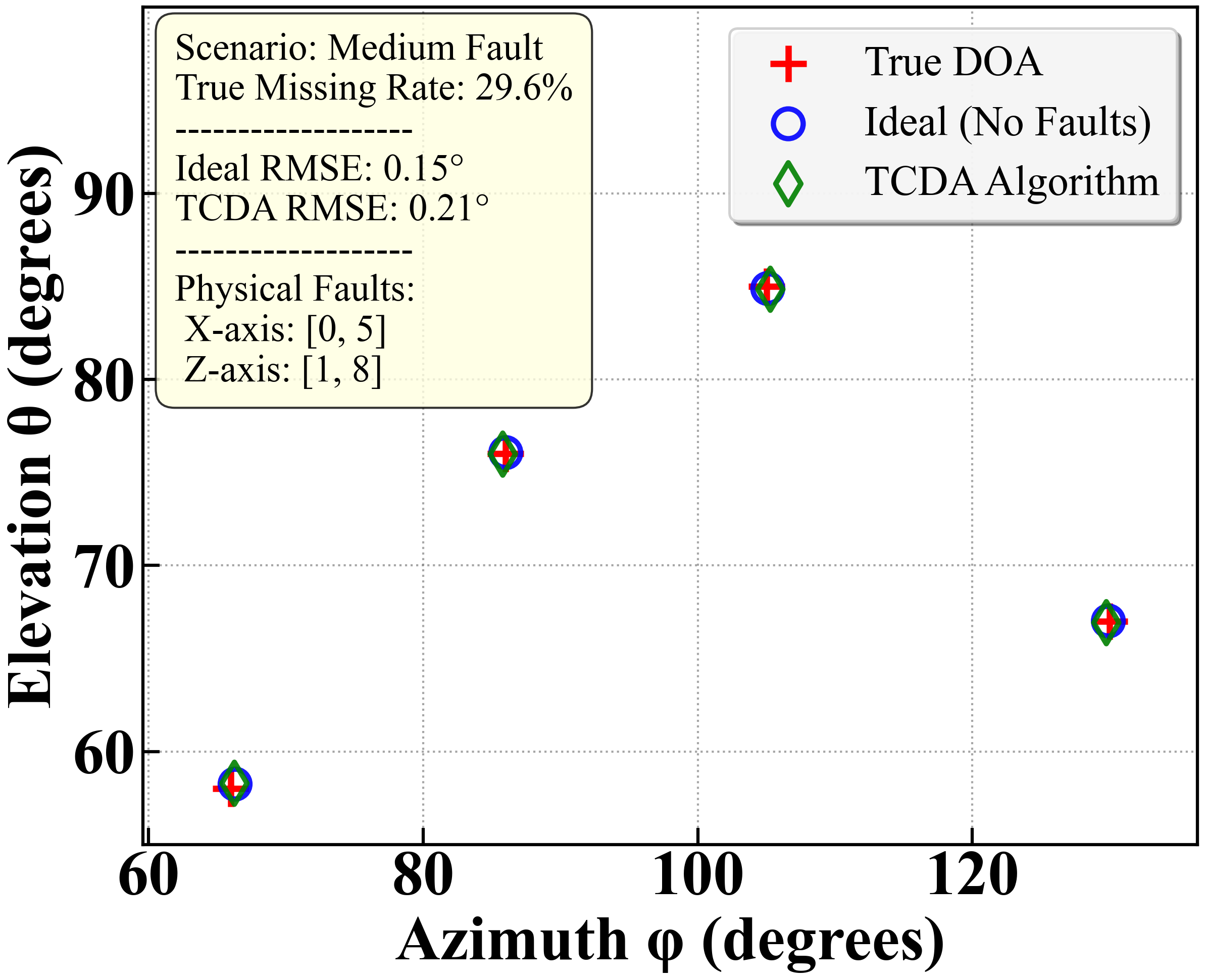}
    \centerline{(b) Medium}\medskip
  \end{minipage}
  
  \vspace{0.3cm} 

  \begin{minipage}[b]{0.48\linewidth}
    \centering
    \includegraphics[width=\linewidth]{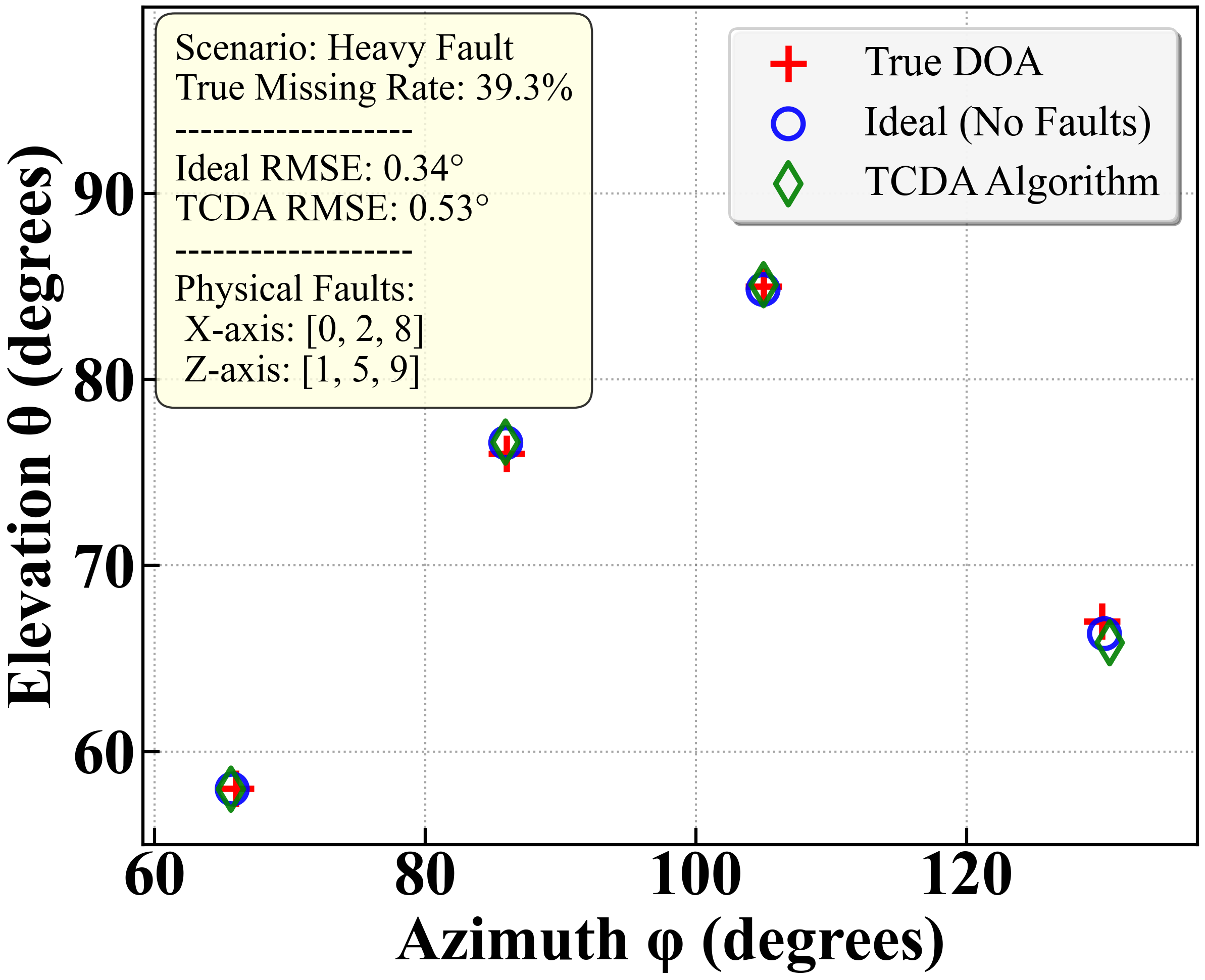}
    \centerline{(c) Heavy}\medskip
  \end{minipage}
  \hfill
  \begin{minipage}[b]{0.48\linewidth}
    \centering
    \includegraphics[width=\linewidth]{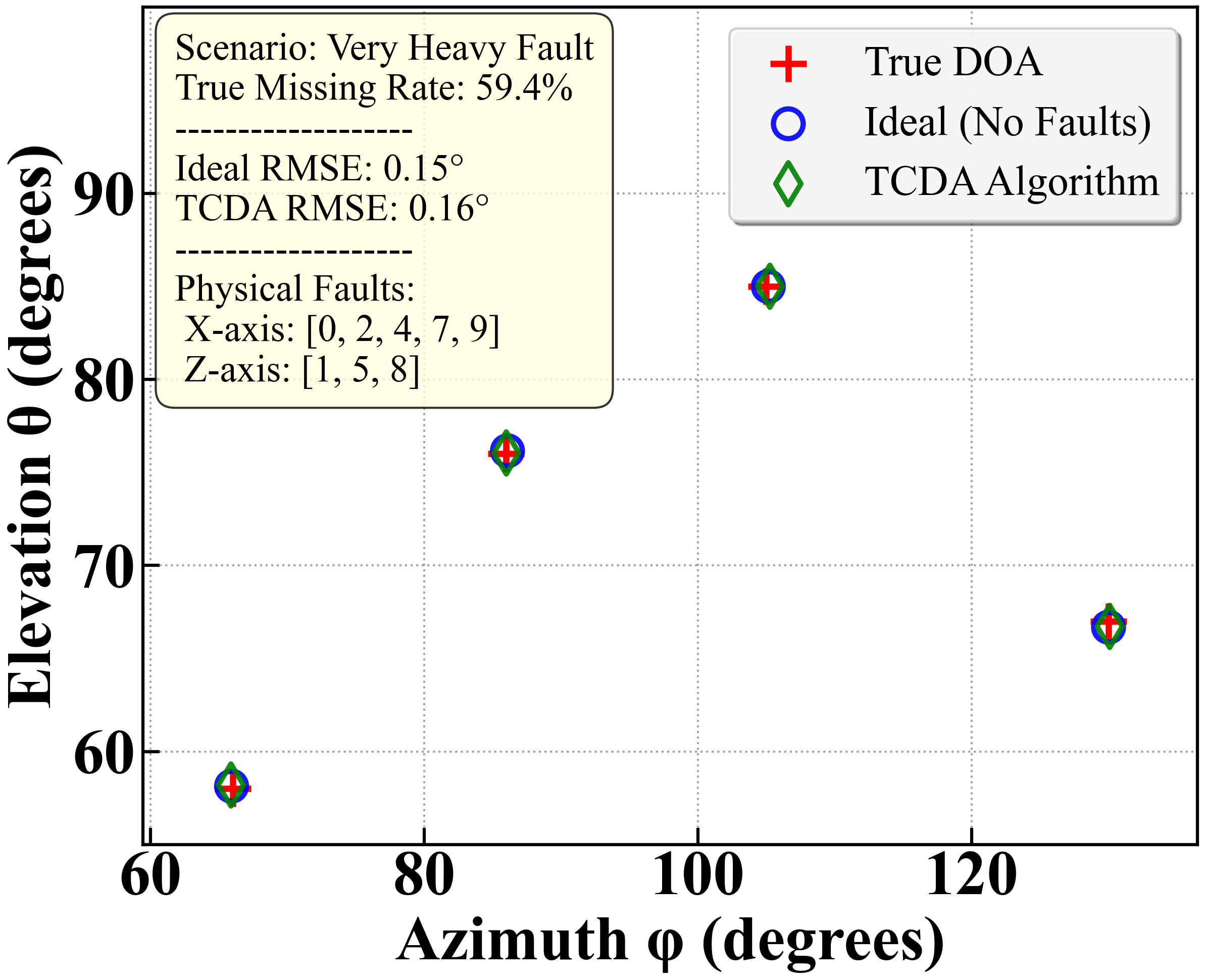}
    \centerline{(d) Very Heavy}\medskip
  \end{minipage}

  \caption{DOA estimation scatter plots under four fault scenarios at SNR = 10 dB.}
  \label{fig:fourfaults}
\end{figure}

\begin{figure}[t]
  \centering
  \includegraphics[width=\linewidth]{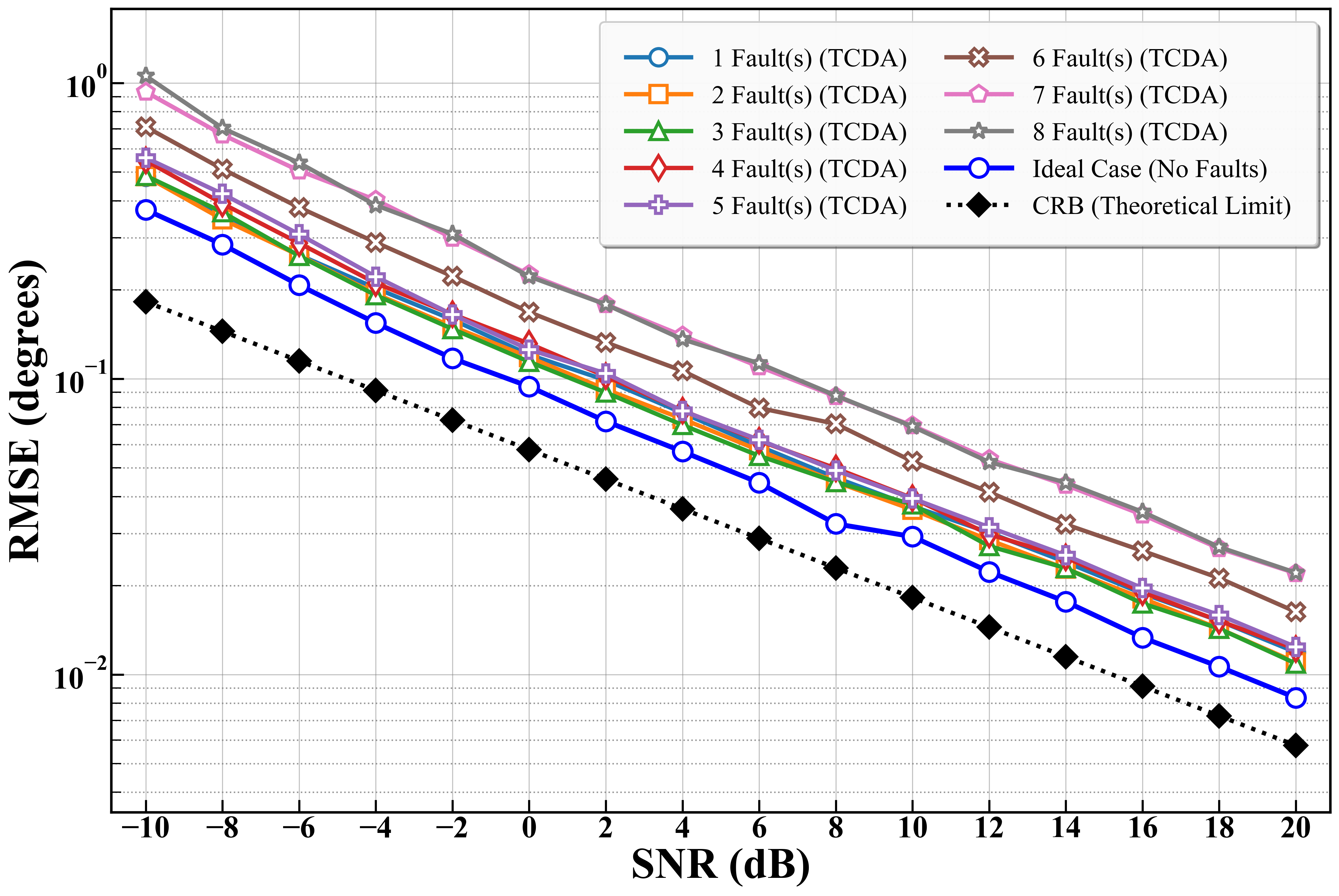}
  \caption{RMSE performance versus SNR for different numbers of faulty sensors.}
  \label{fig:rmse_vs_snr}
\end{figure}
To comprehensively evaluate the performance of the proposed algorithm, a series of simulations were conducted. All experiments are based on an L-shaped array with $M=10$ elements, receiving signals from $K=4$ far-field narrowband sources. The true Directions-of-Arrival (DOAs) of the sources are set to $(58^\circ, 66^\circ)$, $(67^\circ, 130^\circ)$, $(76^\circ, 86^\circ)$, and $(85^\circ, 105^\circ)$. The number of snapshots is $N_{\text{snapshots}}=500$. To ensure statistical reliability, all performance results were obtained by averaging over $1000$ independent Monte Carlo trials.


Fig. \ref{fig:fourfaults} presents the DOA estimation scatter plots for a fixed SNR of $10$ dB across four scenarios of increasing severity, visually demonstrating the algorithm's resilience. Even with moderate physical damage resulting in $9.8\%$ missing data in the target tensor (Fig. \ref{fig:fourfaults} (a)), the TCDA estimates (green diamonds) remain tightly clustered around the true DOAs, with an RMSE of $0.20^\circ$ that is remarkably close to the ideal $0.14^\circ$ bound. As the damage intensifies to a medium fault level, causing a significant $29.6\%$ data loss (Fig. \ref{fig:fourfaults} (b)), the algorithm's robustness persists, maintaining an excellent RMSE of $0.22^\circ$.

This resilience is further tested under a heavy fault scenario (Fig. \ref{fig:fourfaults} (c)), where nearly $40\%$ of the tensor data is unavailable. While a slight increase in estimation variance is visible, TCDA's RMSE of $0.33^\circ$ confirms it still resolves all sources accurately. The framework's true power is showcased in the extreme stress test of a very heavy fault (Fig. \ref{fig:fourfaults} (d)). Here, despite a staggering $59.4\%$ of the data being missing—a condition under which traditional methods would catastrophically fail—our algorithm remains operational, providing usable estimates with an RMSE of $0.39^\circ$. Collectively, these results provide compelling evidence of the TCDA framework's "self-healing" capability. The algorithm's performance degrades gracefully as physical damage increases, effectively bridging the gap between catastrophic failure and the performance of a perfect array, making it a highly practical solution for real-world applications.

To further investigate the performance of the TCDA algorithm under varying noise levels and for different numbers of faults, we conducted a second set of experiments, with the results presented in Fig. \ref{fig:rmse_vs_snr}. The figure plots the RMSE as a function of SNR, ranging from $-10$ dB to $20$ dB, for scenarios with $1$ to $8$ randomly faulty sensors.

As is evident from the figure, all performance curves behave consistently: the RMSE decreases monotonically as the SNR increases, and all curves lie above the Cramér-Rao Bound (CRB). Contrary to typical sub-optimal estimators, the curves do not exhibit distinct "error floors" in the high-SNR region; instead, they maintain a parallel descent relative to the ideal case and the CRB. The performance limits are distinctly categorized by the severity of the lost spatial information. For mild to moderate damage ($1$ to $5$ faulty sensors), the performance curves are tightly clustered and closely follow the curve for the ideal (no-faults) case across the entire SNR spectrum, indicating that the tensor completion effectively compensates for the missing data. In contrast, for severe fault scenarios ($6$ to $8$ faults), the curves clearly separate from the ideal case, reflecting the fundamental performance limit imposed by substantial spatial information loss.

The most crucial insight from this figure is the "graceful degradation" property exhibited by the TCDA algorithm. Performance does not collapse catastrophically after a few faults; instead, it declines smoothly and predictably as the number of faults increases. Notably, with $1$ to $5$ faulty sensors, the algorithm suffers virtually no performance loss compared to the ideal case. Even in the extreme stress test of $8$ faulty sensors, the algorithm maintains robust performance, keeping the RMSE well below $1^\circ$ for any SNR above $-8$ dB (e.g., $0.70^\circ$ at $-8$ dB). This predictable and smooth performance decline confirms that the algorithm is a highly reliable and robust choice for applications facing varying noise conditions and inevitable hardware failures.

\section{CONCLUSION}
This paper introduces the TCDA framework, an innovative solution for robust 2D-DOA estimation with defective sensor arrays. We transform the physical fault problem into a standard weighted PARAFAC decomposition task by modeling the incomplete third-order target tensor from the defective signals. This approach is both theoretically elegant and computationally efficient, as its output factor matrices are directly used for DOA extraction, eliminating extra parameter pairing steps. A detailed derivation of the ALS-based solver is provided. Extensive simulations confirm that TCDA effectively recovers the incomplete spatial data, exhibiting a remarkable "graceful degradation" property. Notably, the algorithm entirely avoids the error floors typical of sub-optimal estimators and maintains estimation accuracy approaching that of a flawless array even when half of the sensors are defective. This research opens new avenues for developing highly reliable, fault-tolerant sensor systems for complex operational environments.

\end{document}